\documentclass[prd,showpacs,amsmath,amssymb,nofootinbib,twocolumn]{revtex4}

\newcommand\lsim{\lesssim}
\newcommand\gsim{\gtrsim}

\renewcommand\({\left(}
\renewcommand\){\right)}
\renewcommand\[{\left[}
\renewcommand\]{\right]}

\newcommand\eq[1]{Eq.~(\ref{#1})}

\newcommand\eqst[2]{Eqs.~(\ref{#1})--(\ref{#2})}

\newcommand\eqreff[1]{(\ref{#1})}

\newcommand\ee{\end{equation}}
\newcommand\be{\begin{equation}}
\newcommand\eea{\end{eqnarray}}
\newcommand\bea{\begin{eqnarray}}

\newcommand\mpl{M_{\rm P}}

\def\calp{{\cal P}}

\newcommand\GeV{\,\mbox{GeV}}

\newcommand\sub[1]{_{\rm #1}}

\newcommand\mone{^{-1}}
\newcommand\mtwo{^{-2}}

\newcommand\mhalf{^{-1/2}}
\newcommand\half{^{1/2}}

\newcommand\quarter{^{1/4}}

\newcommand\calpz{\calp_\zeta}

\newcommand\phic{\phi \sub{end}}

\newcommand{\etaz}{\eta_0}
\newcommand{\meta}{|\eta_0|}

\usepackage{graphicx}
\begin{document}

\title{More hilltop inflation models}

\author{Kazunori Kohri, Chia-Min Lin, David H.~Lyth}

\affiliation{{\it Physics Department, Lancaster University, Lancaster
LA1 4YB, UK}}

\begin{abstract}
Using analytic expressions, we explore the parameter space for hilltop
inflation models with a potential of the form $V_0\pm m^2\phi^2 -a\phi^p$.
With the positive sign and $p>2$ this converts the original hybrid inflation
model into a hilltop model, allowing the spectral index to agree with the
observed value $n=0.95$.
In some  cases the observed value is theoretically favored,
while in others there is only the generic prediction $|n-1|\lsim 1$.

\end{abstract}

\maketitle


\section{Introduction}

The general idea of what has been called \cite{bl} hilltop inflation is that
cosmological scales leave the horizon while the inflaton is near the top of a
hill, with  its potential  still concave-downward. This allows the initial
condition to be set by an era of eternal inflation, whose indefinite duration
may  remove any concern about the probability of arriving at the
hilltop in the first place. Hilltop inflation also ensures that the spectral
tilt is negative, though it may not be guaranteed that the amount of tilt
is small as is required by observation.

It was noticed earlier \cite{bl} that hilltop inflation is more natural than
one might think. Starting with any rather flat potential, it is easy to
generate a maximum with  a reasonable-looking additional term.
In this paper we pursue that line of thinking, by considering a potential
which covers a range of possibilities,  and is yet simple enough to yield
analytic formulas for the predictions.

We shall take for granted the basic ideas of slow-roll inflation
model-building, as explained for instance in \cite{treview,book,al,paris}.
Cosmological scales leave the horizon during about ten $e$-folds of
inflation. The value $\phi$ of the
inflaton field when  $N$ $e$-folds of inflation remain is
\be
N = \mpl\mtwo \int^{\phi(N)}_{{\phic}} \frac{V}{V'} d\phi
, \label{nofphi} \ee
where $\mpl=(8\pi G)\mhalf =2.2\times 10^{18}\GeV$.
Cosmological scales leave the horizon during about ten $e$-folds of inflation,
starting with the largest scale $k=H_0$. (Here $k$ is the present value
of the comoving wavenumber
and $H_0$  the present Hubble parameter).
For a standard cosmology after inflation, the value of $N$ when the latter
scale leaves the horizon is typically in the range $50$ to $60$. In following
we give the predictions for $N=60$.

We assume  that the primordial
curvature perturbation $\zeta$ is generated from the vacuum fluctuation
of the inflaton, instead of later by some curvaton-type mechanism.
The spectrum of the tensor perturbation, as a fraction $r$ of the observed
spectrum of the curvature perturbation, is $r=16\epsilon$, where
 $\epsilon=(\mpl^2/2)(V'/V)^2$. (Here and in the following all functions
of $\phi$ are to be evaluated when the relevant scale leaves the horizon.)
 If the potential
remains concave-downward for the rest of inflation, this implies \cite{bl}
\be
r < 0.002 \(\frac{\Delta\phi}\mpl \)^2 \( \frac{60}N \)^2
, \ee
where   $\Delta\phi$ is the  variation of the inflaton field
We will here demand only that the shape of the potential is such as to give
$r\lsim 10\mtwo$. This is consistent with present observation.

The cmb anisotropy determines  the  magnitude of the spectrum of the
curvature perturbation \cite{wmap3} as $\calp_\zeta=(5\times 10^{-5})^2$,
with an uncertainty which is negligible in the present context.
Invoking the slow-roll prediction for
$\calp_\zeta$ one finds
\be
r= 16\epsilon =  \( \frac{V_0\quarter}{3.3\times 10^{16}\GeV} \)^4
. \ee
We call the second equality the cmb normalization.
Our  requirement $r<0.01$ corresponds to
\be
V_0\quarter<1.0\times 10^{16}\GeV = 4.2\times 10^{-3}\mpl
. \label{upperV} \ee

The spectral index of the curvature perturbation is defined by \be
n-1 = d\calp_\zeta/d\ln k = -d\calp_\zeta/dN , \ee where $N$ is the
number of $e$-folds remaining after the scale $k$ leaves the horizon
and the final equality assumes slow roll. The slow-roll prediction
is \be n=1+2\eta-(3/8)r , \label{etapred} \ee where  $\eta\equiv
\mpl^2V''/V$. Assuming $r\lsim 10\mone$, the cmb anisotropy requires
\cite{wmap3} \be n=0.948^{+0.015}_{-0.018} \label{nobs} . \ee An
analysis including other types of observation \cite{combined} finds
instead $n=0.97\pm 0.01$. When we invoke the observational value we
take the central value of the cmb result.

Now comes an important point.
Since we are assuming  $r\lsim 10\mtwo$, the observed value of $n$
means that we can take the prediction to be simply
 $n=1+2\eta$ if it is to fit observation.  We conclude that
{\em if a slow-roll
model of inflation is to generate the observed curvature perturbation
with $r\lsim 10\mtwo$,
its potential must be concave-downward while cosmological scales leave the
horizon, with $\eta\equiv \mpl^2 V''/V \simeq -0.02$.}

We should also consider the running $n'\equiv dn/\ln k= -dn/dN$.
The models we consider here give $n'>0$ and this condition should be imposed
as a prior when determining the upper bound on $n'$ allowed by
observation. As will be shown elsewhere \cite{klm},  current observations
require in that case something like $n'< 0.01$. We impose this constraint
where relevant.

\section{The potential}

We consider a  potential  of the following form, with $\lambda$
positive.
\bea
V(\phi) &=&  V_0 \pm  \frac12  m^2\phi^2 - \lambda \frac{\phi^p}{\mpl^{p-4}}
+\cdots \\
&\equiv& V_0 \(  1 + \frac12  \etaz  \frac{\phi^2}{\mpl^2} \)
 - \lambda \frac{\phi^p}{\mpl^{p-4}}+\cdots
\label{hilltop}
, \eea
with
\begin{eqnarray}
    \label{eq:eta0_deff}
    \eta_{0} = \frac{\pm m^{2} M_{\rm p}^{2}}{V_{0}}.
\end{eqnarray}
The additional terms are presumed negligible during inflation and we
 require that $\phi$ rolls towards the origin.
There is no need to assume
that $p$ is integral if this is just regarded as a parameterisation of the
potential.

To  justify keeping just two terms, especially if $p$ is an integer,
we would like $\phi\ll\mpl$ after the largest cosmological scale leaves the
horizon (which we are taking to correspond to $N=60$). We will just
 require that the maximum of $\phi\sub{end}$ and $\phi(N=60)$
is  $\lsim \mpl$.

\begin{figure}[htbp]
\begin{center}
\includegraphics[width=0.45\textwidth]{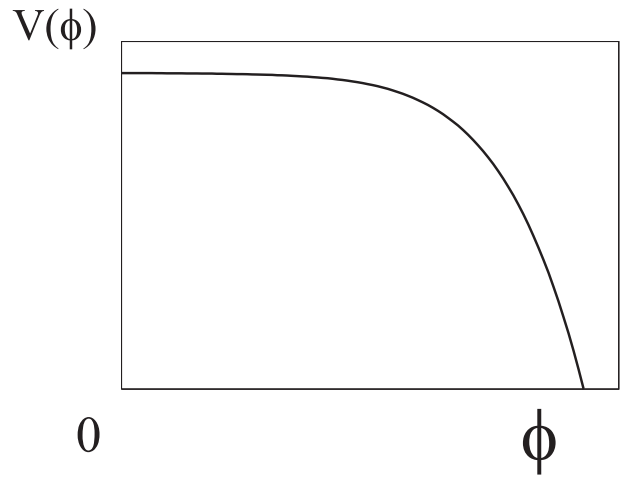}
\caption{\footnotesize{Model1. Two-term approximation about the hilltop.}}
\label{pot1}
\end{center}
\end{figure}

\begin{figure}[htbp]
\begin{center}
\includegraphics[width=0.45\textwidth]{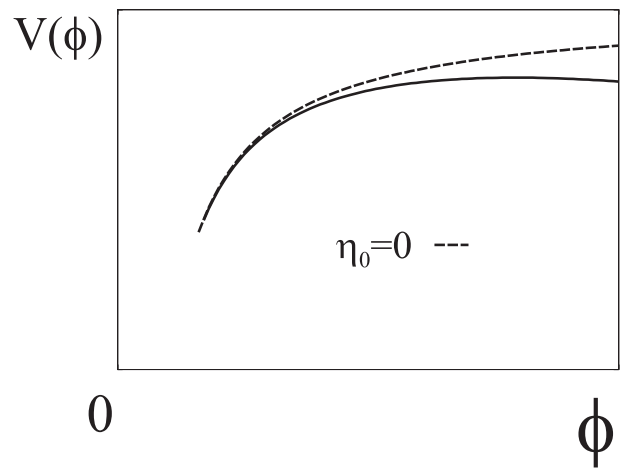}
\caption{\footnotesize{Model2. Hilltop mutated/brane hybrid inflation.}}
\label{pot2}
\end{center}
\end{figure}

\begin{figure}[htbp]
\begin{center}
\includegraphics[width=0.45\textwidth]{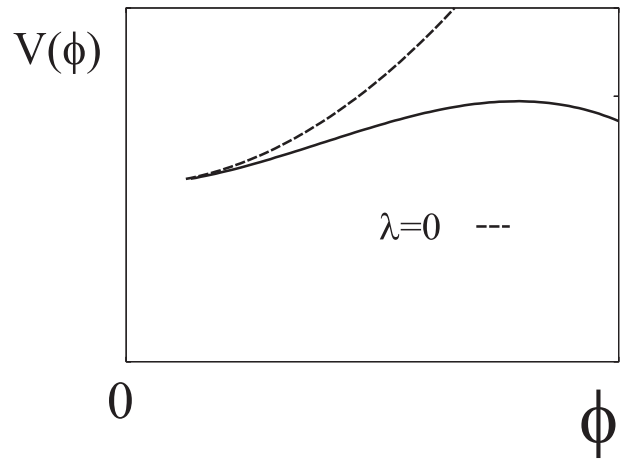}
\caption{\footnotesize{Model3. Hilltop original hybrid inflation.}}
\label{pot3}
\end{center}
\end{figure}

We consider three different
regimes of $\etaz$ and $p$.
\begin{enumerate}
\item \textbf{Two-term approximation about the hilltop.}
The choice  $\etaz\leq 0$ and $p>2$ sets  $\phi=0$ at the hilltop,
and  adds a higher power to the quadratic term.
This is sketched in Figure \ref{pot1}.
\item \textbf {Hilltop mutated/brane  hybrid  inflation.} The choice
 $\etaz<0$ and $p<0$ converts these models to hilltop models,
as sketched in Figure \ref{pot2}.
\item \textbf{ Hilltop tree-level hybrid inflation.}
The choice $\etaz>0$ and $p>2$ converts the original hybrid inflation model
to a hilltop model, as illustrated in Figure \ref{pot3}.
\end{enumerate}

To achieve slow-roll with the potential \eqreff{hilltop}, we need $V\simeq
V_0$ giving
\bea
\frac {V'}V &=& \eta_0 \frac\phi{\mpl^2} -
p\lambda \frac{\phi^{p-1}}{V_{0} \mpl^{p-4} } \\
\frac {V''}V &=& \frac{\eta_0}{M_{\rm p}^{2}}
  - p(p-1) \lambda \frac{\phi^{p-2}}{V_{0}\mpl^{p-4} }
. \eea Then \eq{nofphi} has the analytic solution \cite{grs}
\bea
\( \frac\phi\mpl \)^{p-2} & =& \( \frac{V_0}{\mpl^4} \) \frac{
\eta_0 e^{(p-2) \eta_0 N} } {\eta_0 x  +  p \lambda \(
e^{(p-2)\eta_0 N}-1 \) }
\label{phi} \\
x &\equiv&  \( \frac{V_0}{\mpl^4} \) \( \frac \mpl {\phic} \)^{p-2}
, \eea
leading to the predictions
\bea
\calp_\zeta &=& \frac{1}{12\pi^2}
\(
\frac{V_0}{\mpl^4}
\)^{\frac{p-4}{p-2}} e^{-2\eta_{0}N} \nonumber \\
&\times&
\frac{ \[p\lambda(e^{(p-2) \eta_{0}N}-1)+\eta_0 x\]^\frac{2p-2}{p-2}
}{
\eta_0^\frac{2p-2}{p-2} \( \eta_0 x- p\lambda \)^2 }
\label{cmb}\\
n-1&=&2\eta_{0}\left[1-\frac{\lambda p(p-1)e^{(p-2)\eta_{0}N}}
{\eta_0 x+ p\lambda(e^{(p-2)\eta_{0}N}-1)}\right]
\label{npred} \\
n'&=& 2\eta_0^2\lambda p(p-1)(p-2) \nonumber \\
&\times& \frac{ e^{(p-2)\eta_0 N}    (\eta_0 x- p\lambda) }
{ \[ \eta_0 x +  p\lambda \( e^{ (p-2)\eta_0 N} -1  \) \]^2 }
. \eea

In a supergravity theory where $\meta$ vanishes in the global supersymmetric
limit, the {\em generic} expectation is $|\eta|\sim 1$. For slow-roll inflation
{\em per se}, all we need is  $|\eta_0| \ll 1$ which does not necessarily imply
significant fine-tuning. The problem (usually called the $\eta$ problem)
comes though when one tries to understand why the spectral index
given by \eq{etapred}  is so close to 1. We will keep this issue in mind,
and review the final situation in the Conclusion after considering in turn
each of the three cases.

\section{Two-term approximation}

The two-term approximation may  apply in a  wide variety of cases,
and has a long history. In using our parameterisation we choose the
origin $\phi=0$ to be a maximum of $V$ (at least in the regime
$\phi>0$. The necessary condition $V'(0)=0$ might be  ensured by a
symmetry (with the origin  the fixed point of the transformation)
but that is not essential. The  maximum of the potential   might be
occur because the potential is periodic (corresponding to $\phi$
being a pseudo Nambu-Goldstone boson (PNGB)) or else through the
interplay of two terms as in our cases 2 and 3.

We will focus on the non-hybrid case, where $V(\phi)$ descends
smoothly to a minimum with  $V=0$. Then we might be dealing with
modular inflation corresponding to $\Delta\phi\sim\mpl$. (`Moduli'
allowing this kind of inflation may be expected in string theory as
discussed for instance in \cite{racetrack}.) Alternatively we might
be dealing with a  small-field model, corresponding to $\Delta\phi$
some orders of magnitude below $\mpl$. If the origin is a fixed
point of symmetries involving $\phi$ we deal with what  one might
call new inflation.  (The original new inflation model \cite{new}
(see also \cite{hm}) corresponds more or less to $p=4$ and
$\eta_0=0$ which we consider below.)


This case has been investigated in \cite{grs,grslett}, where the formulas
 \eqst{phi}{npred} were first given.\footnote
{These authors had in mind the case of new inflation.}
Less complete investigations were made   earlier,
keeping just the $\phi^p$ term  \cite{treview,al,paris}, both terms with
 $p=4$ \cite{dineriotto,treview,pseudonatural}, and both terms with generic
$p$ \cite{supernew}.) In the following we present a
further investigation of this interesting case.

\subsection{Spectral index}

The  potential is concave-downward throughout inflation. As a
parameterization over a limited range it makes sense to allow $p$ to
be non-integral, with  $p\gsim 3$.

Since $V'/V$ is increasing during inflation, one expects
the value of $\phi(N)$
given by \eq{nofphi} to be insensitive to $\phi\sub{end}$, at least some
part of the parameter space. That would correspond to $x=0$ being a good
approximation.
We will  not make that approximation,
 but instead take $\phic$ (the point at
which slow-roll fails) to be the point where $\eta(N=0)=-1$ which
corresponds to $n(N=0)=-1$ or
\be
\frac{V_0}{\mpl^4} =  \frac{ p(p-1)\lambda }{1-\eta_0}
\(\frac\phic\mpl \)^{p-2}
. \label{naivex} \ee
Since slow-roll requires $\meta\ll 1$, our requirement $\phic\lsim \mpl$
becomes
\be
V_0/\mpl^4\lsim \lambda
. \label{vbound1} \ee

Using \eq{naivex} we
find that the spectral index depends only on $p$ and $\etaz$.
For $\etaz=0$ we recover the known result \cite{treview}
\be
    n= 1 - 2\(\frac{p-1}{p-2}\)\frac1 N
. \label{knownn}     \ee
This is automatically within  the observed range.

For $\lambda=0$ we recover the known result
\be
n-1=2\eta_0
. \label{known2} \ee
 In that case,
the requirement $\phi\sub{end}< \mpl$ requires $\meta\gsim 1$
corresponding to $n-1\simeq -1$. The value of $\meta$
is in mild conflict with the slow-roll requirement, and of course
the value of $n$ is far below the observed one.

\begin{figure}[htbp]
\begin{center}
\includegraphics[width=0.50\textwidth]{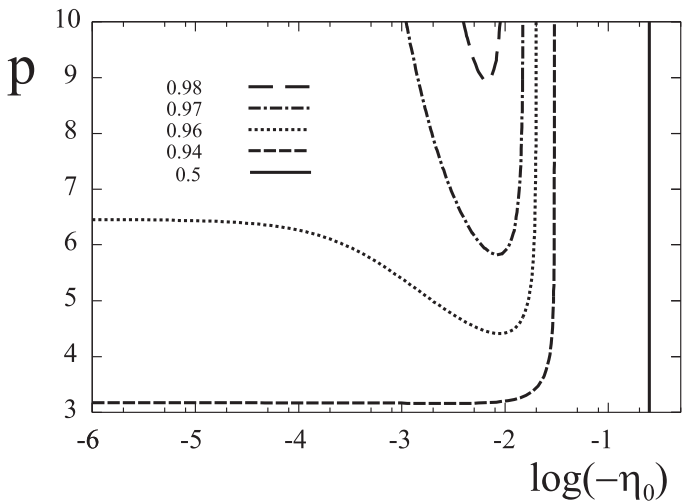}
\caption{\footnotesize{$p\geqslant3$, contours of $n$ in the $\log(-\etaz)$-$p$ plane.}}
\label{pot4}
\end{center}
\end{figure}

In Figure ~\ref{pot4} we plot contours of constant $n$ in the
 $p$-$\etaz$ plane. The nearly horizontal lines correspond to the
limit \eqreff{knownn}. The nearly vertical lines correspond to the
limit \eqreff{known2} being practically attained at $\phi(N=60)$, the
term of V proportional to $\phi^p$ ensuring  that slow-roll ends
at $\phi\lsim \mpl$.

\begin{figure}[htbp]
\begin{center}
\includegraphics[width=0.50\textwidth]{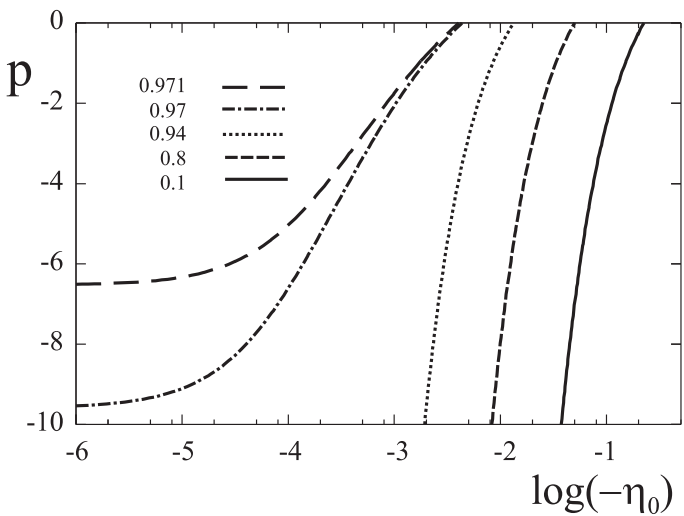}
\caption{\footnotesize{$p<0$, contours of $n$ in the
$\log(-\etaz)$-$p$ plane.}}
\label{fig4}
\end{center}
\end{figure}

\subsection{Modular inflation}

 We now look at the cmb normalization, beginning with modular inflation.
Here the minimum of $V$ is expected to be of order $\mpl$, corresponding
 roughly to $\lambda = V_0/\mpl^4$. We will impose that equality, leaving
only the parameters $\eta_0$ and $p$.

\begin{figure}[htbp]
\begin{center}
\includegraphics[width=0.50\textwidth]{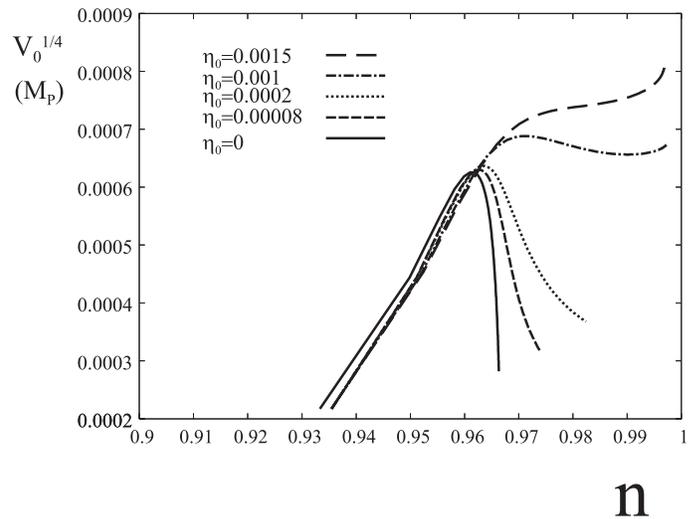}
\caption{\footnotesize{Model1, $p=3\sim100$, different values of $\etaz$ in $V_0^{1/4}$-$n$ plane. We use $\lambda=V_0/\mpl^4$ which corresponds to modular inflation.}}
\label{d0528}
\end{center}
\end{figure}

As  $\phi$ is not far below $\mpl$ we should view \eq{hilltop} just as an
approximation, valid hopefully for some $p\gsim 3$.
In Figure  \ref{d0528} we plot the cmb-normalized $V_0\quarter$ against
$n$, for a few values of $\eta_0$ and the range $3<p<100$. This plot shows if we demand a high inflation scale in a modular inflation, then $n$ in a modular inflation model cannot be far below $1$.

\subsection{New inflation}

 For new inflation corresponding to $\phi\sub{end}\ll \mpl$, it
 is reasonable to  suppose
 that the term $\propto \phi^p$  is the leading next term in a
power-series expansion, further terms being suppressed by the
small value of $\phi$. Then $p$ will be  an integer bigger than 2.
The integer can be bigger than 3 because in
 new inflation the origin is supposed to be a fixed point of
symmetries.

\subsubsection{Case $p=4$}

Let us suppose that odd $p$ are  forbidden by a symmetry $\phi\to-\phi$,
making the  leading term  $p=4$.
This gives   the original new inflation model \cite{new}, except that the
slight running of $\lambda$ with $\phi$ invoked there is absent.
A famously small value of $\lambda$ is required.
The  original version of the model, where $\lambda$ is a gauge coupling, was
therefore rejected but it was soon pointed out \cite{sv}
that $\lambda$ could instead
be a Yukawa coupling making the small value perhaps acceptable. More recently
it has been noticed \cite{ss}
that the predicted spectral index is compatible with
observation.

In the version of the last paragraph this model is not supersymmetric so
that there is no natural expectation that  $\meta$ will be significant,
 but still we are free to consider that case.  Moreover, this potential
with significant $\meta$ has been motivated in other ways, both
non-supersymmetric \cite{pseudonatural} and supersymmetric \cite{dineriotto}.
The  second case however very fine-tuned  \cite{treview}. It therefore appears
that {\em
the case $p=4$ is best realised without supersymmetry, especially if
$|\eta_0|$ is included.}

\begin{figure}[htbp]
\begin{center}
\includegraphics[width=0.50\textwidth]{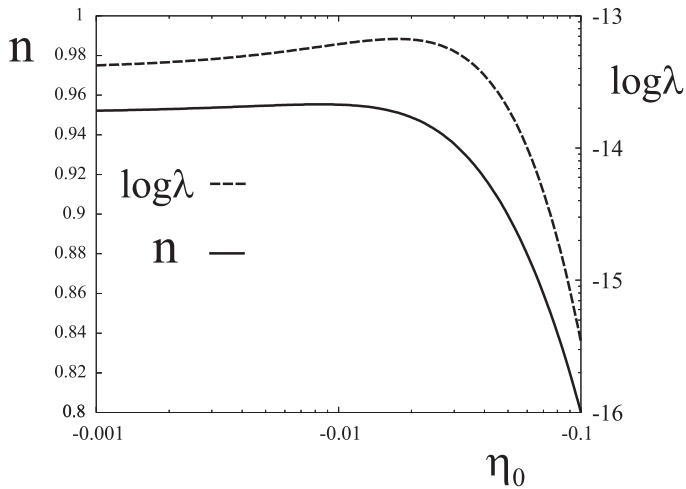}
\caption{\footnotesize{Model 1, $p$=4, $n$ and $\log\lambda$ versus $\eta_0$.}}
\label{pot5}
\end{center}
\end{figure}

After imposing the cmb normalization there is one  parameter
 which we take to be
$\eta_0$. In Figure \ref{pot5} we show $n$ and $\log\lambda$ as functions
of $\eta_0$.
We see in Figure \ref{pot5} that increasing
$|\eta_0|$ decreases both $\lambda$ and $n$, allowing the latter to be well
 below the observed value.

\subsubsection{Case $p=6$}

The need for the quartic term to be very suppressed  is a generic
feature of inflation models after the cmb normalization is imposed
\cite{treview}.  Supposing it to be negligible and still
taking $V$ to be even, we expect the leading term to be $p=6$.

\begin{figure}[htbp]
\begin{center}
\includegraphics[width=0.50\textwidth]{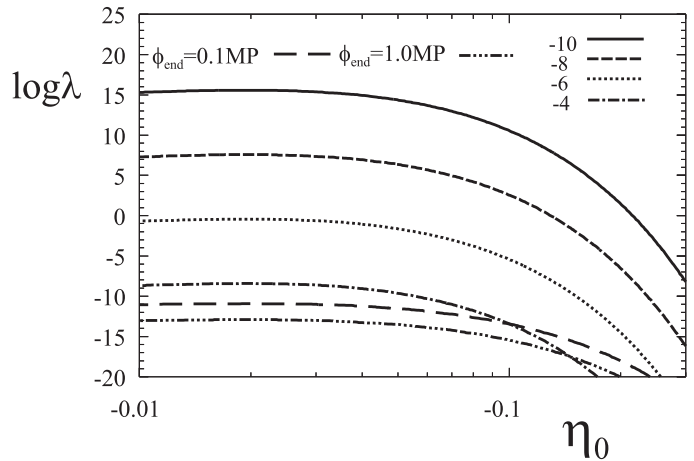}
\caption{\footnotesize{Model 1, $p$=6, contours of
$\log(V_0^{1/4}/\mpl)$ in the $\eta_0$-$\log\lambda$ plane. We also
show the lines $\phi_{\rm end}=0.1\mpl$ and $\phi_{\rm end}=1.0\mpl$.}}
\label{pot6}
\end{center}
\end{figure}

\begin{figure}[htbp]
\begin{center}
\includegraphics[width=0.45\textwidth]{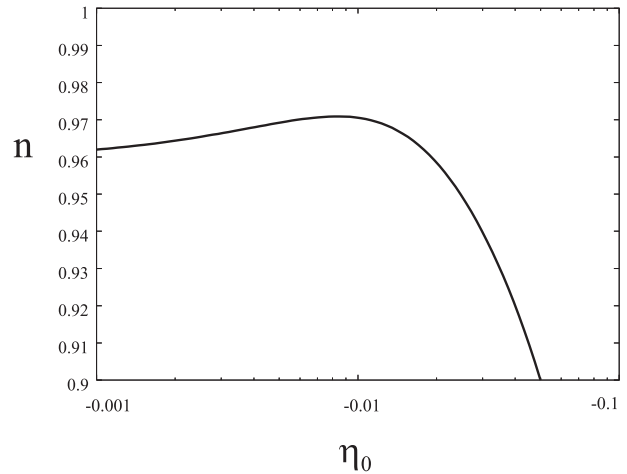}
\caption{\footnotesize{Model 1, $p$=6, $n$ versus $\eta_0$.}}
\label{pot7}
\end{center}
\end{figure}

\begin{figure}[htbp]
\begin{center}
\includegraphics[width=0.45\textwidth]{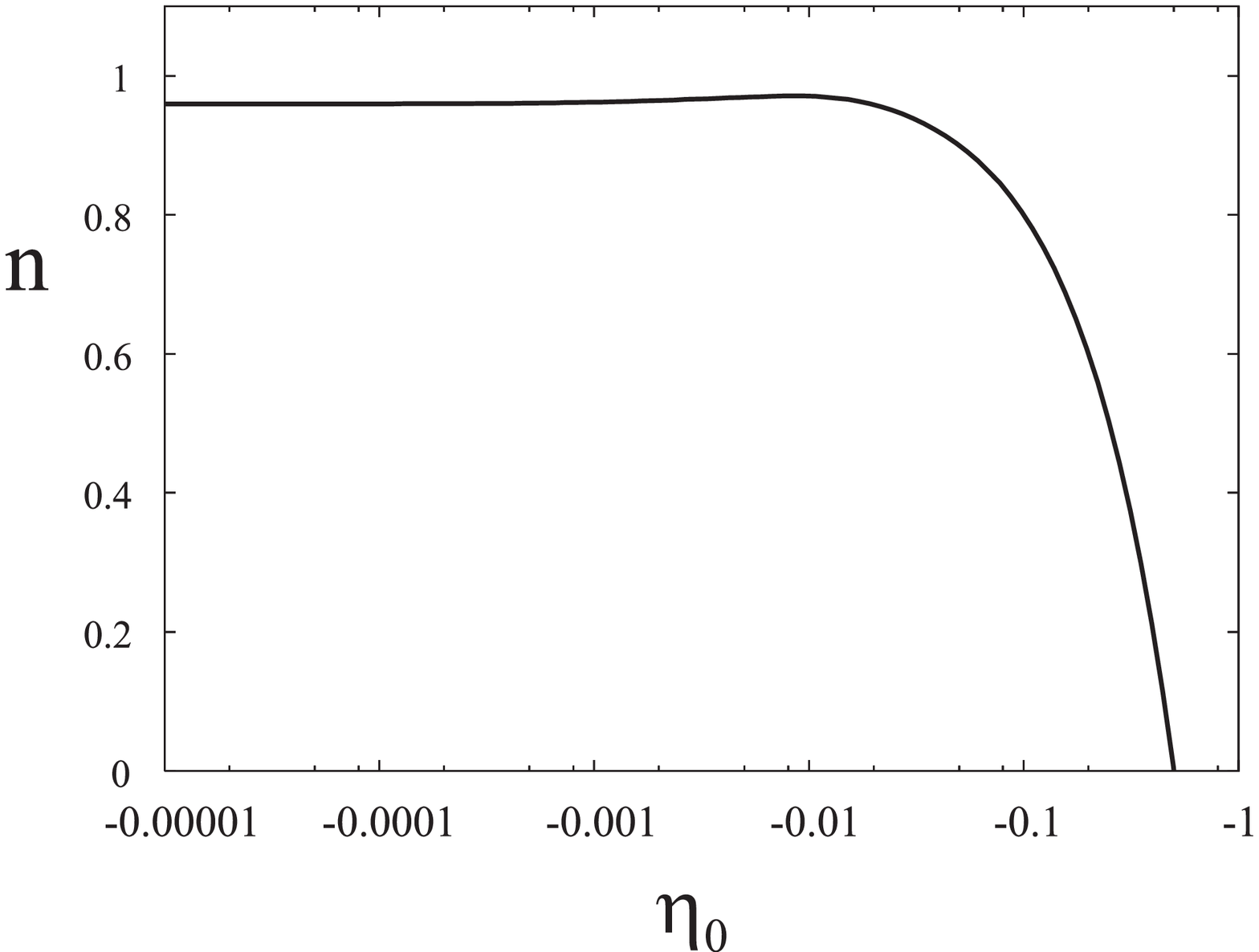}
\caption{\footnotesize{Model 1, $p$=6, $n$ versus $\eta_0$. This plot shows we can have a big range of $n$. The limit of $n$ when $\etaz\rightarrow0$ is $n=0.959$ .}}
\label{fig7}
\end{center}
\end{figure}

In Figure \ref{pot6}, we show contours of $\log (V_0^{1/4}/\mpl)$
in the $\etaz$-$\log\lambda$  plane, assuming that $x$ is given by
\eq{naivex}. We also show the lines $\phic=0.1\mpl$ and $\phic=1.0\mpl$.
In Figures \ref{pot7} and \ref{fig7} we show $n$ against $\eta_0$.
From Figure \ref{fig7} , we can see the spectral index can be far below the
observed value.

To suppress the quartic term we would
like to impose supersymmetry. With a minimal Kahler potential we arrive
at the $\eta_0=0$ case with a superpotential of the form
 \cite{supernew}
\be
W = V_0\half \Psi \[ 1- \(\frac \Phi v \)^{p/2} \] + \cdots
\label{supers}
, \ee
where the extra terms ensure $\Psi=0$ without affecting
the potential in the direction of the inflaton $\phi\equiv |\Phi|$.


\section{Hilltop mutated hybrid inflation}

Now we come to the second case, characterized by $\etaz<0$ and $p<0$.
The prediction for $n$ is given in Figure \ref{fig4}. Regarding the
limit of large $p$   and small $\meta$, the remarks made for the previous
case apply.

This case corresponds to what is called mutated hybrid inflation
\cite{mut,smooth,ourmut}.  It differs from ordinary hybrid inflation
in the following respect. Ordinary hybrid inflation assumes a coupling
between the inflaton field $\phi$ and the waterfall field $\chi$
which is some function of $\phi$ times $\chi^2$. This fixed $\chi$
at the origin during slow-roll inflation, which
ends
when $\phi$ passes through some critical value which destabilizes $\chi$.
In mutated hybrid inflation, the coupling involves a higher power than
$\chi^2$. As a result,  the  waterfall field  is not fixed, but instead
 adjusts  to  minimize the potential
as the inflaton field slowly rolls. The inflationary potential is
$V(\phi,\chi(\phi))$ and inflation ends when slow-roll fails.

Until now, it has usually been assumed that $V$ with  $\chi=0$ is absolutely
flat. The most general potential  that has been considered \cite{ourmut} is
then
\be
V=V_0 - A\chi^s + B \chi^q\phi^r
\label{mut} . \ee
When the potential is minimized by $\chi(\phi)$ at fixed $\phi$,
the potential becomes
\be
V(\phi,\chi(\phi))= V_0 -
\[
 A\(\frac{Bq}{As}\)^\frac s{s-q}
-B\(\frac{Bq}{As}\)^\frac q{s-q} \]
\phi^\frac{-rs}{q-s}
. \ee

This is of the form \eq{hilltop} with $\eta_0=0$ and $p = -(sr)/(q-s)$.
If $\phi=0$ is a fixed point of the symmetries, it
is quite reasonable to add a term $\pm\frac12m^2\phi^2$ to the above
potential, to arrive at \eq{hilltop}. More generally though,
a linear term would be allowed in the potential and so it would be
more reasonable to add a term $\pm \frac12m^2(\phi-\phi_0)^2$
with $\phi_0$ a parameter.

The latter is the case for the original mutated hybrid inflation model
\cite{mut}. It is not covered by our parameterisation, and we
 focus instead  on
what was called smooth hybrid inflation \cite{smooth}.
There  the origin
is a fixed point, with
\be
V = \(V_0\half - \frac{\chi^4}{16\mpl^2} \)^2
+ \frac{\chi^6 \phi^2}{16\mpl^4}
. \ee
During inflation we can drop the $\chi^8$ term, to get
\eq{mut} with  $s=4$,  $q=6$ and $r=2$. This leads
\eq{hilltop} with $p=-4$ and
\be
\lambda=(2/27) (V_0\quarter/\mpl)^6
, \ee
and of course $\etaz=0$.
We can easily compare this with \cite{smooth}.
First, we use equation (11), and take the limit $\eta_0\rightarrow0$ for
$p=-4$. We obtain $n=1-5/180\simeq0.97$, this is exactly the same as
\cite{smooth}. Second, we can see from (10) and use the relation
$V_0=m^2\mpl^2/\eta_0$, we obtain
$V_0/\mpl^4\simeq1.12\times10^{-8}\lambda^{1/4}$. Use the above relation
$\lambda=(2/27) (V_0\quarter/\mpl)^6$ we can solve for
$(V_0^{1/4}/\mpl)\simeq5.08\times10^{-4}$.

\begin{figure}[htbp]
\begin{center}
\includegraphics[width=0.50\textwidth]{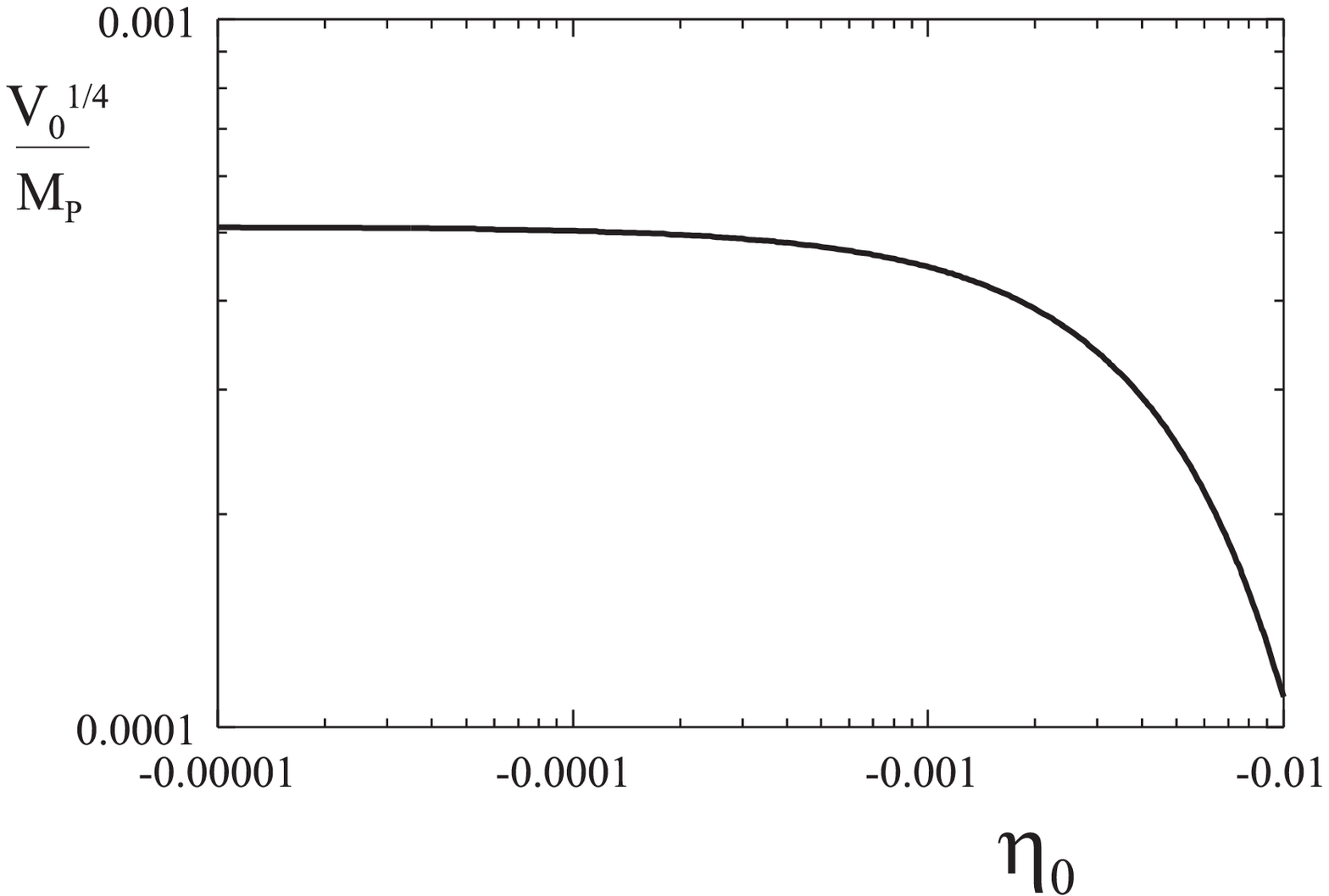}
\caption{\footnotesize{$p$=-4, $V_0^{1/4}/\mpl$ versus $\eta_0$.}}
\label{0510}
\end{center}
\end{figure}

\begin{figure}[htbp]
\begin{center}
\includegraphics[width=0.50\textwidth]{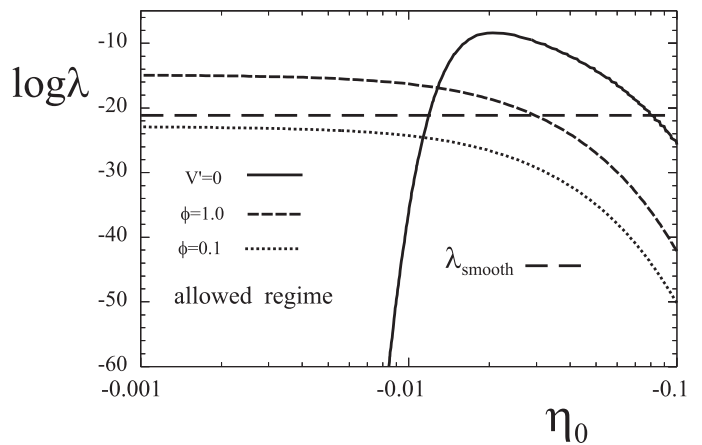}
\caption{\footnotesize{Model 2, $p$=-4, We show the lines $V'=0$,
$\phi=1.0\mpl$, and $\phi=0.1\mpl$ in the $\eta_0$-$\log\lambda$
plane. Below the line $V'=0$, the inflaton rolls to the right. The
allowed region in our model is the left lower corner in the
plot. $\lambda_{\rm smooth}=1.27\times10^{-21}$ correspond
to smooth hybrid inflation.}}
\label{vp2}
\end{center}
\end{figure}

The inclusion of  a no-zero $\etaz$ for smooth hybrid inflation
has been  considered in \cite{rss} and we
 further investigate it now. In Figure \ref{vp2} we see that $|\eta_0|$
cannot be very big. {}From Figures \ref{0510} and \ref{potx1} we see that
this requires a high inflation scale, and a spectral index around
the observed value.

The case $p=-4$ has also been derived in a  colliding brane
scenario \cite{collbrane}.  Our potential does not apply in that case,
because  the origin will not be  a fixed point
of symmetries so that a linear term in $\phi$ is allowed which would
go beyond our parameterisation.
(Also, non-canonical normalization is allowed in this case, though
there is a significant regime of parameter space where the normalization is
canonical.)

\begin{figure}[htbp]
\begin{center}
    \includegraphics[width=0.50\textwidth]{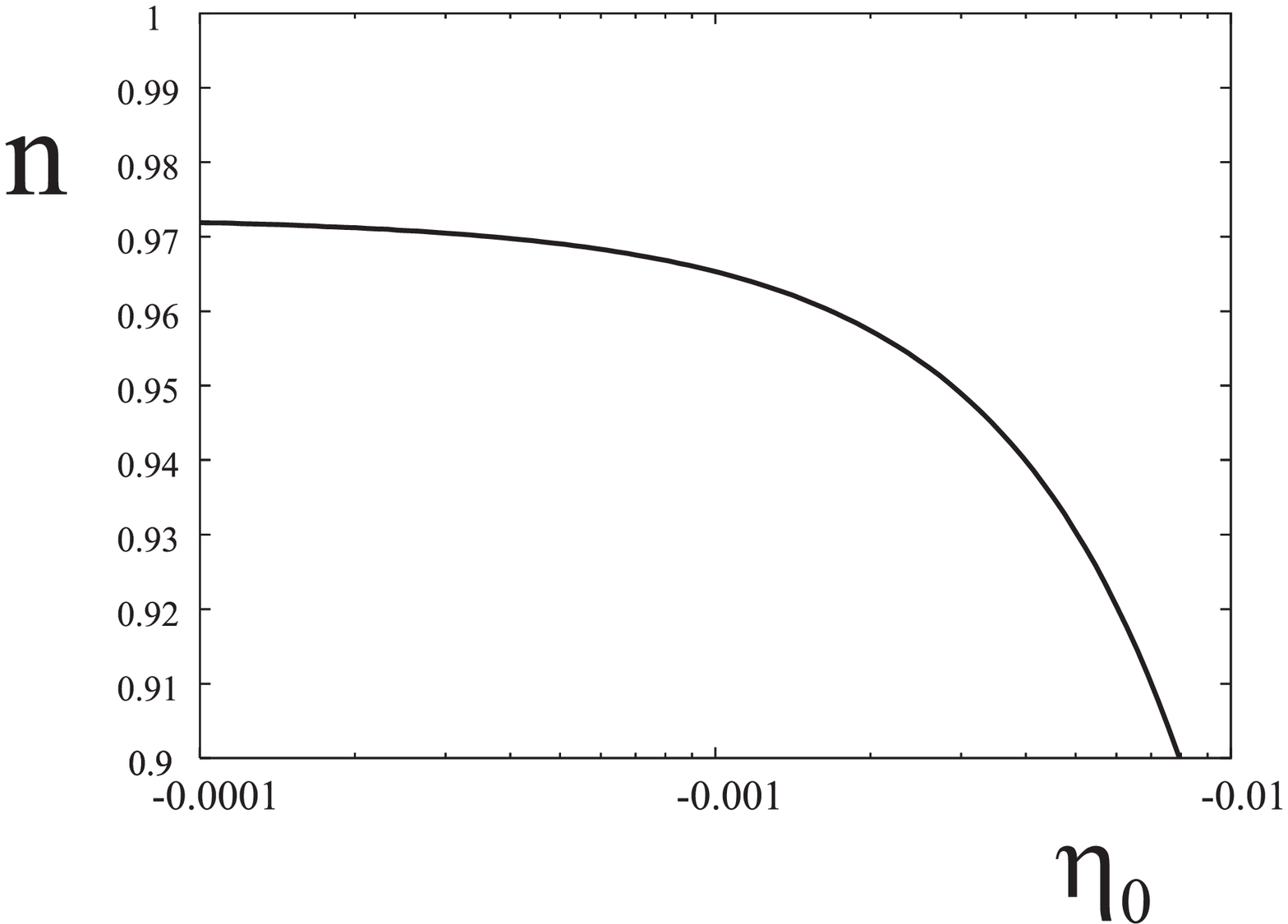}
\caption{\footnotesize{Model 2, $p$=-4, $n$ versus $\eta_0$.}}
\label{pot10}
\end{center}
\end{figure}

\begin{figure}[htbp]
\begin{center}
\includegraphics[width=0.50\textwidth]{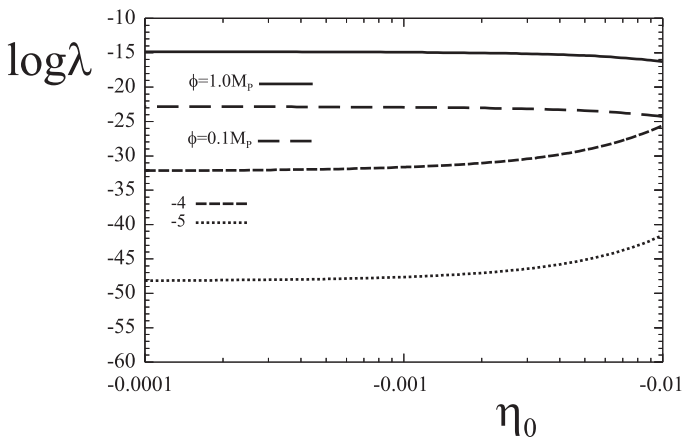}
\caption{\footnotesize{Model 2, $p$=-4, contours of
$\log(V_0^{1/4}/\mpl)$ in $\etaz$-$\log\lambda$ plane.
We require $\phi<\mpl$. }}
\label{potx1}
\end{center}
\end{figure}

\begin{figure}[htbp]
\begin{center}
\includegraphics[width=0.50\textwidth]{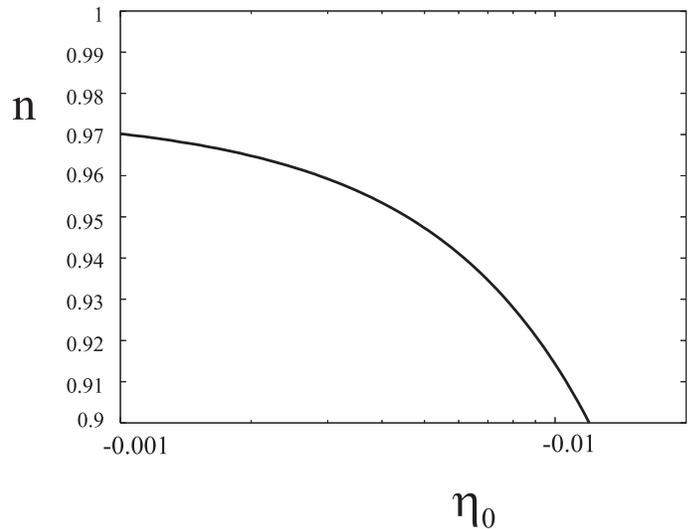}
\caption{\footnotesize{Model 2, $p$=-2, $n$ versus $\eta_0$.}}
\label{pot8}
\end{center}
\end{figure}

\begin{figure}[htbp]
\begin{center}
\includegraphics[width=0.45\textwidth]{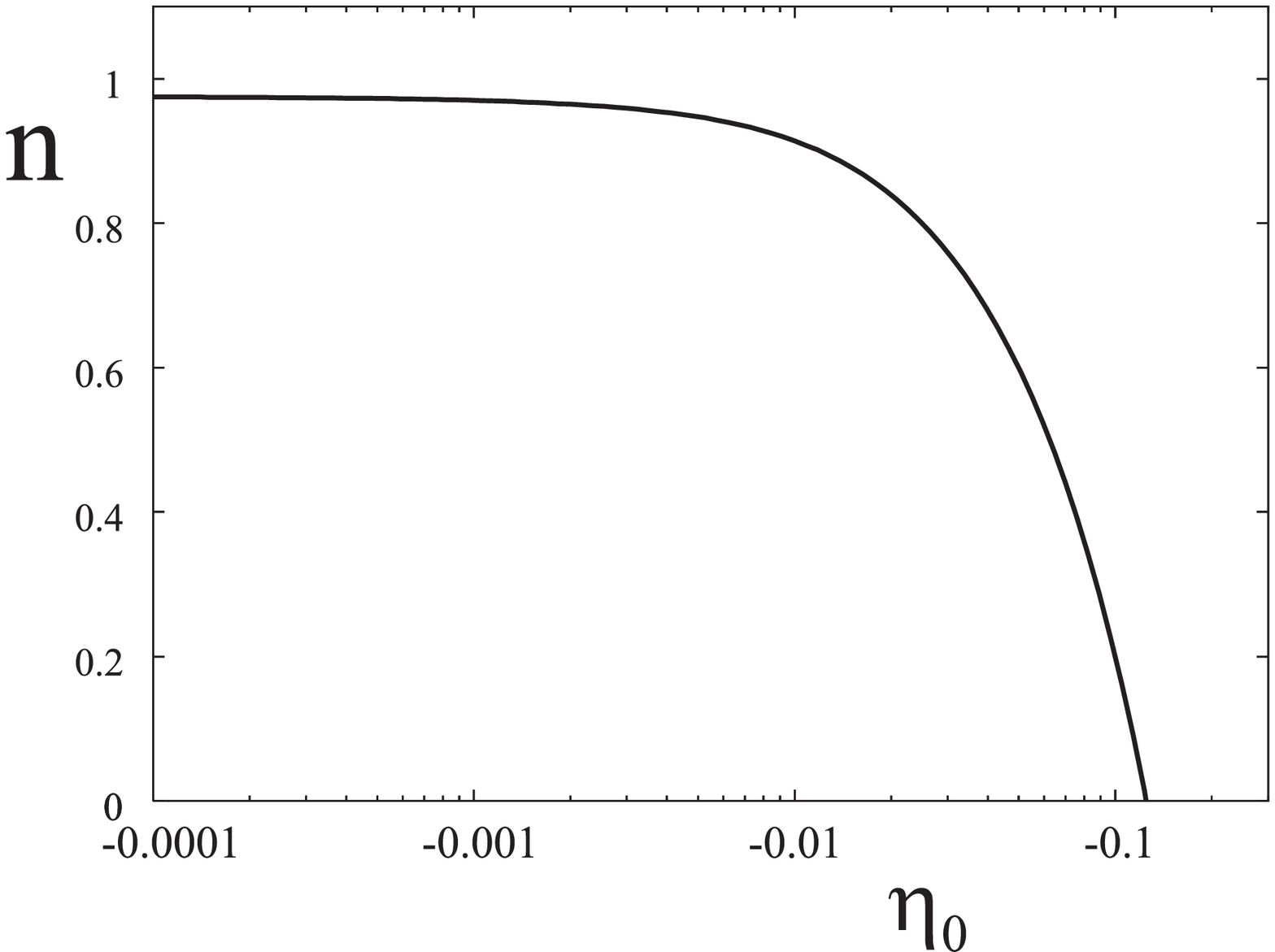}
\caption{\footnotesize{Model 2, $p$=-2, $n$ versus $\eta_0$. This plot
shows we can have a big range of $n$. The limit of $n$ when $\eta_0
\rightarrow0$ is $n=0.975$ .}}
\label{fig08}
\end{center}
\end{figure}

We also considered mutated hybrid inflation
with  $p=-2$, assuming that the origin is a fixed point of the symmetries.
 We plot the spectral index $n$ against $\etaz$ in Figs \ref{pot8} and \ref{fig08} for $p=-2$.
These plots show that basically we can have a big range of spectral index $n$ by introducing nonzero $\etaz$ in our models.

\begin{figure}[htbp]
\begin{center}
\includegraphics[width=0.50\textwidth]{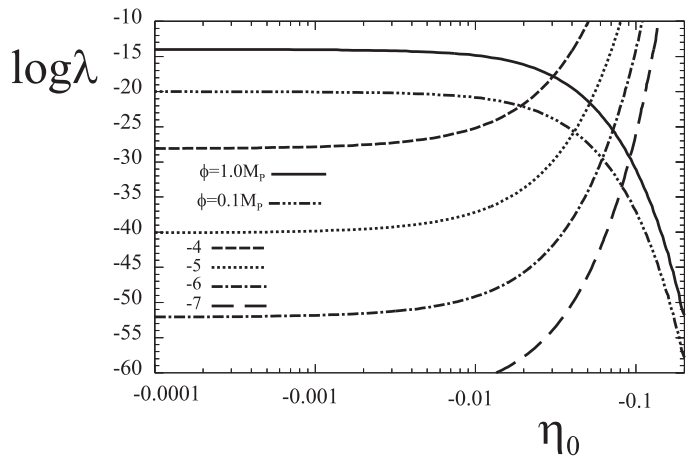}
\caption{\footnotesize{Model 2, $p$=-2, contours of
$\log(V_0^{1/4}/\mpl)$ in $\etaz$-$\log\lambda$ plane. We require $\phi<\mpl$.}}
\label{pot9}
\end{center}
\end{figure}


\section{$F$- and $D$-term inflation}

Now  we  consider  the standard $F$- and $D$-term
inflation models \cite{fterm,dterm}.
  Here, {\em sticking to the
simplest versions of those models}, $\lambda\phi^p /\mpl^{p-4}$ is
replaced by $V_0 (g^2/4\pi^2) \log(\phi/Q)$ with $Q$ a constant of
order $\phi$.  Since we are assuming $V\simeq V_0$,
other,
this  is equivalent to take the limit $p\to 0$ with $\lambda p$
fixed at the value
\begin{eqnarray}
    \label{eq:lambdaP}
   \lambda p = - \frac{V_0}{\mpl^4} \frac{g^2}{4\pi^2}.
\end{eqnarray}
This case has been analyzed for both positive \cite{panag,mydterm,km}
and \cite{bl,john}  negative  $\etaz$,
but the latter choice leading to hilltop inflation is more
interesting and makes it easier for the models to agree with observation
and we analyze it further now.

\begin{figure}[htbp]
\begin{center}
\includegraphics[width=0.50\textwidth]{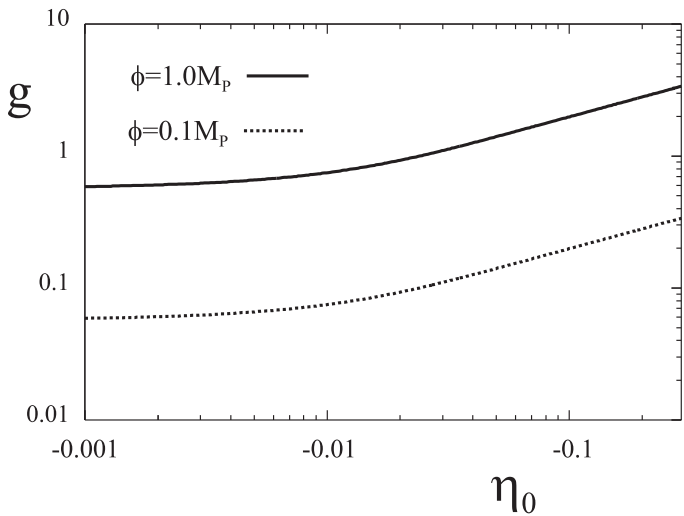}
\caption{\footnotesize{
This plot shows contours of $\phi$ in the $\etaz$-$g$ plane.}}
\label{v_prime}
\end{center}
\end{figure}

We find
\be
\phi(N=60)=\frac{g}{2\pi}\frac{[\frac{-\etaz}{1+\etaz}-(1-e^{-120\etaz})]^{1/2}}{(-\etaz)^{1/2} e^{-60\etaz}}
.  \ee
which is independent of $V_0$. This is shown in
 Figure  \ref{v_prime}. We see that the requirement
$\phi\ll\mpl$ requires $g\ll 1$.

\begin{figure}[htbp]
\begin{center}
\includegraphics[width=0.50\textwidth]{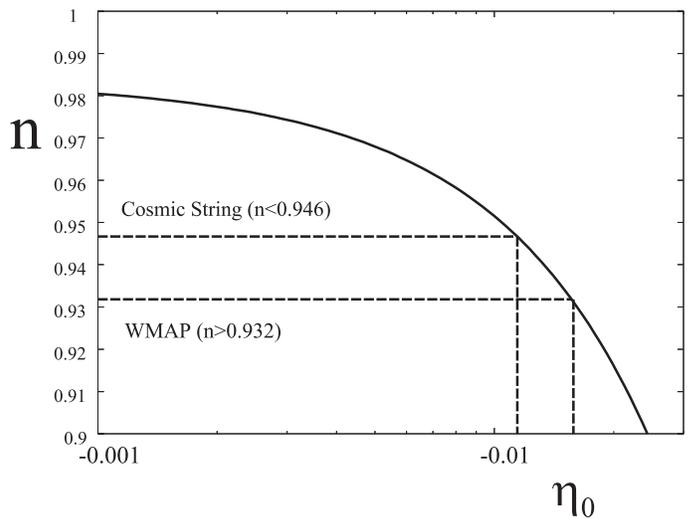}
\caption{\footnotesize{$p$=0, $n$ versus $\eta_0$. The upper bound from
cosmic strings applies to the  $D$-term case
 \cite{john}.}}
\label{pot43}
\end{center}
\end{figure}

The spectral index is
\be
n=1+2\etaz\left[1-\frac{e^{-120\etaz}}{1-e^{-120\etaz}+\frac{\etaz}{1+\etaz}}\right]
,  \ee
 shown in Figure \ref{pot43}.

\begin{figure}[htbp]
\begin{center}
\includegraphics[width=0.50\textwidth]{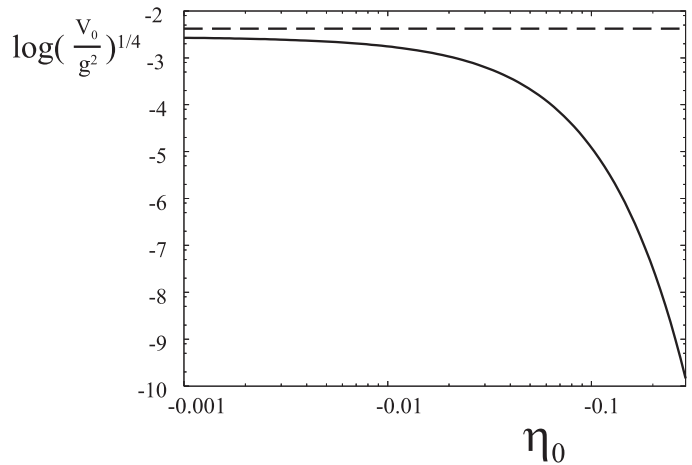}
\caption{\footnotesize{$p$=0, $(V_0/g^2)^{1/4}$ versus $\eta_0$ in
Plank unit $\mpl=1$. We also show the requirement as in \eq{upperV} in the dashed line which correspond an upper bound for $g=1$.}}
\label{d0518}
\end{center}
\end{figure}

The  cmb normalization is
\be
\frac{V_0\quarter}{\sqrt g}=\left(\frac{2.5\times10^{-5}\sqrt{12}(-\etaz)^{1/2}(1+\frac{\etaz}{1+\etaz})}{e^{-60\etaz}[\frac{-\etaz}{1+\etaz}-(1-e^{-120\etaz})]^{1/2}}\right)^{1/2}
.  \ee
This is
plotted in Figure  \ref{d0518}. In both the $F$- and $D$-term models,
$V_0\quarter/\sqrt g$ is the vev of the waterfall field. In the $F$-term
case one supposes that the waterfall field is a GUT Higgs field making
its vev of order the GUT scale $10^{16}\GeV$, and a similar value is
expected in the $D$-term case with a high string scale. Imposing that
restriction, we see from Figure  \ref{pot43} that $n$ cannot be far
below the observed value.


\section{Hilltop tree-level hybrid inflation}

\subsection{Generic case}

The third case has never been considered before.
Here we are seeing whether
the original hybrid inflation model  \cite{andreihybrid},
which gives $n\geq 1$ in contradiction
with observations, can be saved by the addition of the $\phi^p$ term,
as in in Figure \ref{pot3}.

 In this case we  keep $x$ as a parameter of
the model.
The value of $x$ depends on $\phic=\phi\sub c$. If the coupling of the
inflaton to the waterfall field is $\lambda_{\phi\chi}\chi^2\phi^q/\mpl^{q-2}$,
then $\phi\sub c^q = \lambda_{\phi\chi}^{-1} m_\chi^2\mpl^{q-2}$ (where
$m_\chi$ is the tachyonic mass of the waterfall field. This gives
\be
\eta_0 x=(\frac{m}{\mpl})^2 \( \frac{\lambda_{\phi\chi}
\mpl^2}{m_\chi^2} \)^{\frac{p-2}{q}}
. \ee
The plausible range of $x$ is clearly very large.

\begin{figure}[htbp]
\begin{center}
\includegraphics[width=0.50\textwidth]{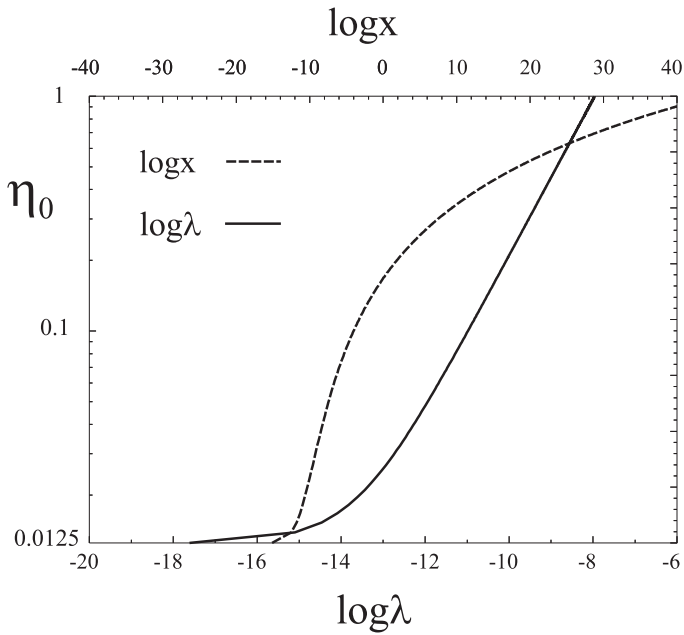}
\caption{\footnotesize{Model 3, $p$=4, $\log x$ and $\log\lambda$
versus  $\eta_0$.}}
\label{pot12}
\end{center}
\end{figure}

\begin{figure}[htbp]
\begin{center}
\includegraphics[width=0.50\textwidth]{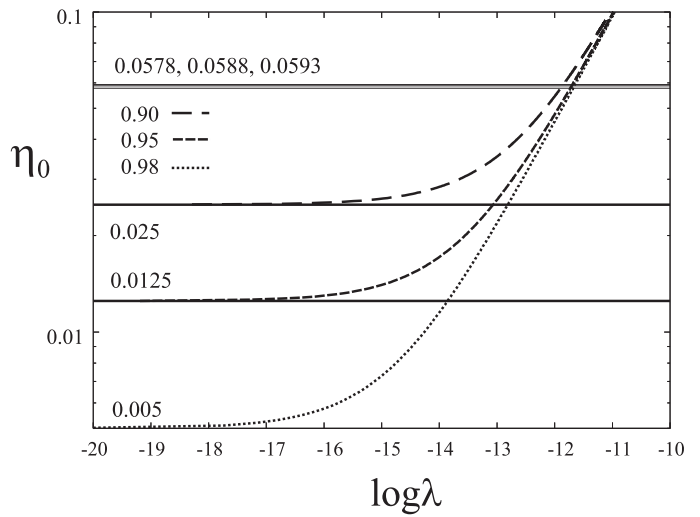}
\caption{\footnotesize{Model 3, $p$=4, contours of n in the
$\log\lambda$-$\eta_0$ plane with the constraint $n=0.9$:
$0.025<\eta_0<0.0578$, $n=0.95$: $0.0125<\eta_0<0.0588$, $n=0.98$:
$0.005<\eta_0<0.0593$. The upper bound is provided by $|n'|<0.01$ and
the lower bound is provided by $\phi_{\rm end}<\phi$}}
\label{fig12}
\end{center}
\end{figure}

To obtain a general idea about the allowed parameter space, we shall
fix $n=0.95$ (the central observed value). Consider first the case
$p=4$.  In Figure \ref{pot12}  we  show $\log x$ and $\log\lambda$ as
a function of $\eta_0$.  At $\eta_0=0.0125$,  $\lambda=x=0$, which is
also the line $\phi=\phi_{\rm end}$ above which we have
$\phi>\phi_{\rm end}$. The curve of $\lambda$ against $\etaz$ also
represent the  contour of $n=0.95$ in the $\lambda$-$\etaz$ plane. We
can generalize this for $n=0.9, 0.95, 0.98$ which we plot in Figure
\ref{fig12}. In Figure  \ref{fig12}, we also show the upper bound for
$\etaz$, which is provided by  $n'<0.01$.

\begin{figure}[htbp]
\begin{center}
    \includegraphics[width=0.50\textwidth]{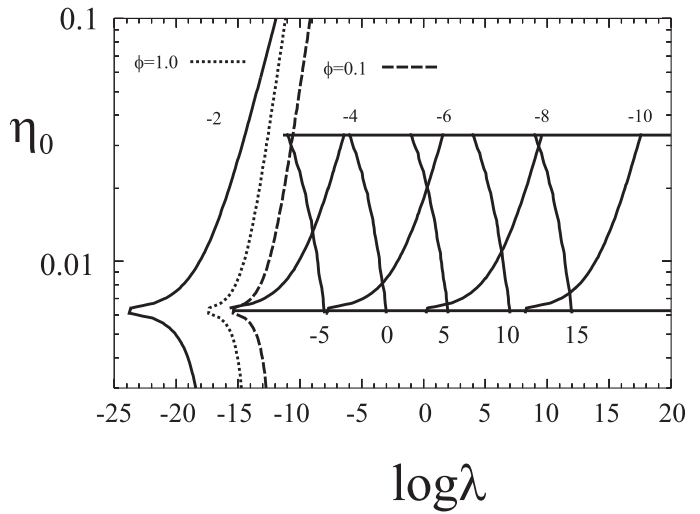}
\caption{\footnotesize{Model 3, $p$=6, contours of
$\log(V_0^{1/4}/\mpl)=-2,-4,-6,-8,-10$, and $\log x=-5,0,5,10,15$. The
dashed line (dotted line) denotes $\phi=0.1\mpl$ ($1\mpl$). The top
horizontal line ($\etaz=0.033$) represents the upper bound on
$\eta_{0}$ by  a condition $|n'|<0.01$. The bottom horizontal line
($\etaz=0.06$) represents the lower bound by a  requirement
$\phi_{\rm end}<\phi$.}}
\label{pot13}
\end{center}
\end{figure}

Now consider $p=6$.  In Figure  \ref{pot13} we show in the
$\eta_0$-$\log\lambda$ plane contours of  $\log x$ and contours of
$\log(V_0\quarter/\mpl)$. We also show the lines  $\phi(N)=\mpl$ and
$\phi(N)=0.1\mpl$. The trapezium-like area (on the r.h.s.  of
$\phi=0.1$ (which means $\phi<0.1$) and $0.006<\eta_0<0.033$)
represents  the allowed region for the inflation model. The upper
horizontal line  $\etaz=0.033$ represents $n'=0.01$, underneath which
is the requirement  $|n'|<0.01$. The lower horizontal line is
$\phi(N)=\phi_{\rm end}$, above which is  the requirement
$\phi(N)>\phi_{\rm end}$ corresponding to motion towards $\phi=0$.  Using
equation (10) and (11), we can see at $\etaz=0.006$, $x=6\lambda$
which  makes $m=0$ therefore $V_0=0$ and $\phi=\phi_{\rm end}=0$ for all
finite  $\lambda$. That is why the plot of $V_0$ and $\phi$ behaves
odd there.


In Figure \ref {pot12}, we see that $\lambda$ has to be very small, in
accordance with the known generic result. On the assumption that it
will actually be negligible after suppression by (say) a supersymmetry
mechanism, we choose instead $p=6$, to obtain the plot shown  in
Figure \ref{pot13}.

The term in the potential proportional to $\phi^p$ may be regarded as
parameterizing physics beyond the ultra-violet cutoff. Taking the
cutoff to be $\mpl$, one might generically expect $\lambda\sim 1$ in
$\lambda \phi^{p}/M_{p}^{p-4}$.  With a lower cutoff $M$ one might
expect $\lambda' \sim (M/\mpl)^{p-4}$ in $\lambda' \phi^{p}/M^{p-4}$
(equivalent to $\lambda\sim 1$ with the replacement $\mpl\to M$.

Within the context of supersymmetry, a simple realization of
hybrid inflation has been termed  \cite{supernatural}
supernatural inflation. Here, the superpotential provides only the coupling
between the waterfall field and the inflaton. All other terms in the
potential, including $V_0$ come from soft supersymmetry breaking.
As in the generic case, our addition of the term proportional to
$\phi^p$ can rescue the model by  allowing it to give a spectral index
in agreement with observation.

\subsection{Black hole formation}

As sketched in Figure \ref{pot3}, the potential starts out concave-downward
as required by observation, but then turns up. As a result it can be
much flatter at the end of inflation than when cosmological scales leave
the horizon. This allows  $\calp_\zeta$ at the end of inflation to be much
bigger than the observed value and the question arises whether it can be
 of order $10^{-2}$ or so,  leading to the production of black holes.

\begin{figure}[htbp]
\begin{center}
\includegraphics[width=0.50\textwidth]{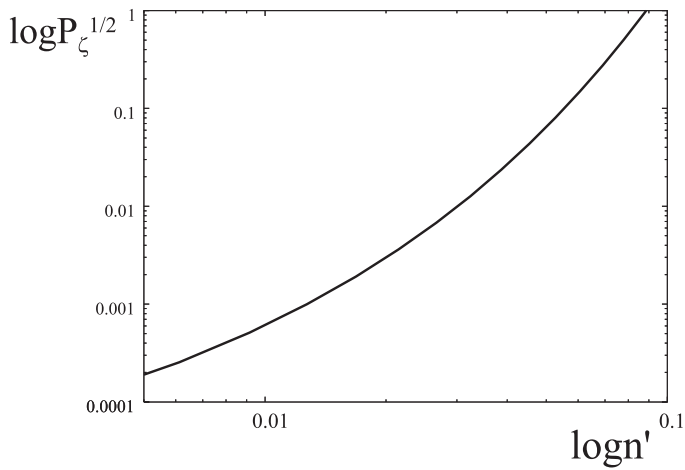}
\caption{\footnotesize{Model 3, $p$=4, $P_{\zeta}^{1/2}(N=0)$ versus $n'$. The upper bound $n'<0.01$ corresponds to $P_{\zeta}^{1/2}(N=0)<6\times10^{-4}$.}}
\label{d0612}
\end{center}
\end{figure}

\begin{figure}[htbp]
\begin{center}
\includegraphics[width=0.50\textwidth]{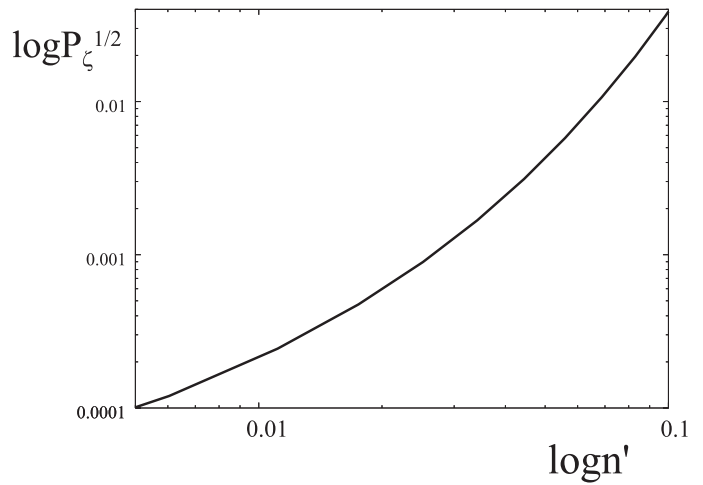}
\caption{\footnotesize{Model 3, $p$=6, $P_{\zeta}^{1/2}(N=0)$ versus $n'$. The upper bound $n'<0.01$ corresponds to $P_{\zeta}^{1/2}(N=0)<2\times10^{-4}$.}}
\label{d061203}
\end{center}
\end{figure}

After fixing $n=0.95$ and apply the cmb normalization, for both the cases
$p=4$ and $p=6$, we can express $P_{\zeta}^{1/2}(N=0)$ and $n'$ as a function
of $\etaz$. We show the plots of $P_{\zeta}^{1/2}(N=0)$ versus $n'$ for the
case $p=4$ and $p=6$ in Figure \ref{d0612} and \ref{d061203}. It is seen that
black hole formation would require  running $n'\sim 0.1$, far in excess of
what is allowed by observation.

It is clear that any potential allowing black hole formation will have a shape
like the one in Figure \ref{pot3}. In  a companion paper \cite{klm}
it  is shown that suitable potentials definitely exist.
A well-motivated example that  seems  viable  at present is the
running-mass model  \cite{running}. With
 the parameters used for Figure 2 of \cite{lgl}, black hole formation
is possible with $n'=0.009$. By altering the gauge group it should be
possible to achieve black hole formation with a lower $n'$.

\section{Conclusion}

We have extended the investigation of \cite{bl}, demonstrating that hilltop
inflation is an absolutely generic possibility for both hybrid and non-hybrid
inflation models. This is a welcome development, in that it allows  a whole
range of models to fit observation while being at the same time well-motivated
from the particle physics viewpoint. The development is disturbing though,
in that it muddies the clean classification of the models presented in for
instance \cite{treview} which could formerly be made on the basis of the
sign and likely value  of the spectral index. However, discrimination between
the models will still be possible if the running can be measured with an
accuracy $\Delta n'\sim 10^{-3}$ (ie.\ an order of magnitude better than
the  bound \cite{klm} provided by present data).

We have explored the parameter space of each model after imposing
the cmb normalization $\calpz^{1/2}= 5\times 10^{-5}$ on the
spectrum. Regarding the spectral index, we have exhibited the effect
of imposing the observed value $n=0.95$, but we have also asked what
range would have been allowed theoretically. The latter question is
reasonable, because in contrast with the normalization of the
spectrum there does not seem to be any anthropic constraint on the
spectral index. Therefore, one would like a value in the right
ball-park to be an automatic consequence of imposing the cmb
normalization.

This desirable state of affairs seems to be achieved for the case of
smooth hybrid inflation and $F$-term inflation, when we require that
these models be part of a GUT theory. It also seems to be the case
for $D$-term inflation and modular inflation, if we require a high
inflation scale corresponding to a high string scale.

In other cases, including  new inflation, the spectral index might
have taken any value in the range $0\lsim 1-n \lsim 1$ demanded by
slow-roll. Tree-level hybrid inflation even allows the whole range
$-1\lsim 1-n \lsim 1$. It was pointed out some time ago \cite{bdl}
that the same is true for $A$-term inflation \cite{aterm}, where one
deals with a potential that is well-approximated by \be V(\phi)
=V'\phi + \frac16 V'''\phi^3 . \label{aterm} \ee
(The same kind of potential has been found recently \cite{recentmod} in the
context of colliding brane inflation.)
With this potential, $V$
has a maximum in about half of the parameter space, giving a hilltop
model. Therefore, if hilltop inflation is favored on the ground that
 eternal inflation can provide the initial condition these models
automatically give $n<1$ but they do  not automatically place $n$ close to 1.

There is a proposal \cite{racetrack} even in these cases, for understanding why
$n$ is so close to 1.  This is to demand as much slow-roll inflation
as possible, so that the inflated volume created by slow-roll is as large
as possible.   Indeed this demand  will drive  $n$ as close
 to 1 as is allowed by  the parameter space. For new inflation and
 modular inflation we have argued that the demand will make
  $1-n$
of order a few divided by $N$ placing it in the right ball-park.
For tree-level hybrid inflation and the potential \eqreff{aterm} the
demand will instead drive $n$ to be indistinguishable from 1.
In any case, it is not clear to us why one should want
to maximise the amount of slow-roll inflation, 
if there has already been a much larger amount of eternal inflation.

{\it Acknowledgments.}~
The research is supported by  PPARC grants
PPA/G/S/2003/00076 and PP/D000394/1   and by EU grants
MRTN-CT-2004-503369 and MRTN-CT-2006-035863. DHL thanks Liam McAllister for
valuable correspondence about colliding-brane inflation.


\begin{thebibliography}{99}

\bibitem{bl}
 L.~Boubekeur and D.~H.~Lyth,
  JCAP {\bf 0507}, 010 (2005)

\bibitem{treview}
 D.~H.~Lyth and A.~Riotto,
  Phys.\ Rept.\  {\bf 314}, 1 (1999)


\bibitem{book}
A.~R.~Liddle and D.~H.~Lyth, \emph{Cosmological Inflation and
Large Scale Structure}, (CUP, Cambridge, 2000)

\bibitem{al}
 L.~Alabidi and D.~H.~Lyth,
  JCAP {\bf 0605}, 016 (2006).

\bibitem{paris}
  D.~H.~Lyth,
  arXiv:hep-th/0702128.

\bibitem{wmap3}
 D.~N.~Spergel {\it et al.},
  arXiv:astro-ph/0603449.

\bibitem{combined}
 J.~Lesgourgues, M.~Viel, M.~G.~Haehnelt and R.~Massey,
  arXiv:0705.0533 [astro-ph].

\bibitem{klm}
K. Kohri, D.H. Lyth and A. Melchiorri (2007) in preparation.

\bibitem{grs}
 G.~German, G.~G.~Ross and S.~Sarkar,
  Nucl.\ Phys.\  B {\bf 608}, 423 (2001)
  [arXiv:hep-ph/0103243].

\bibitem{grslett}
  G.~German, G.~G.~Ross and S.~Sarkar,
  Phys.\ Lett.\  B {\bf 469}, 46 (1999)

\bibitem{racetrack}
  J.~J.~Blanco-Pillado {\it et al.},
  JHEP {\bf 0609}, 002 (2006)
  [arXiv:hep-th/0603129].

\bibitem{new}
A.~D.~Linde,
Phys.\ Lett.\ B {\bf 108}, 389 (1982);
A.~Albrecht and P.~J.~Steinhardt,
Phys.\ Rev.\ Lett.\  {\bf 48}, 1220 (1982).


\bibitem{sv}
 Q.~Shafi and A.~Vilenkin,
  Phys.\ Rev.\ Lett.\  {\bf 52} (1984) 691.

\bibitem{ss}
Q.~Shafi and V.~N.~Senoguz,
  Phys.\ Rev.\  D {\bf 73} (2006) 127301
  [arXiv:astro-ph/0603830].

\bibitem{hm}
 Hawking, S. W., and Moss, I. G., 1982,
Phys. Lett. B {\bf 110}, 35.

\bibitem{fterm}
c E.~J.~Copeland, A.~R.~Liddle, D.~H.~Lyth, E.~D.~Stewart and D.~Wands,
  Phys.\ Rev.\ D {\bf 49} (1994) 6410;
 E.~D.~Stewart,
  Phys.\ Rev.\ D {\bf 51} (1995) 6847;
 G.~R.~Dvali, Q.~Shafi and R.~K.~Schaefer,
  Phys.\ Rev.\ Lett.\  {\bf 73}, 1886 (1994).

\bibitem{dterm}
 E.~D.~Stewart,
Phys.\ Rev.\ D {\bf 51}, 6847 (1995);
 P.~Binetruy and G.~R.~Dvali,
Phys.\ Lett.\ B {\bf 388} (1996) 241;
E.~Halyo,
Phys.\ Lett.\ B {\bf 387} (1996) 43.


\bibitem{panag}
 C.~Panagiotakopoulos,
Phys.\ Lett.\ B {\bf 402} (1997) 257;
C.~Panagiotakopoulos,
Phys.\ Rev.\ D {\bf 55} (1997) 7335.

\bibitem{mydterm}
 D.~H.~Lyth,
Phys.\ Lett.\ B {\bf 419} (1998) 57.

\bibitem{km}
 C.~F.~Kolda and J.~March-Russell,
Phys.\ Rev.\ D {\bf 60} (1999) 023504.

\bibitem{john}
 C.~M.~Lin and J.~McDonald,
  Phys.\ Rev.\ D {\bf 74}, 063510 (2006).

\bibitem{mut}
  E.~D.~Stewart,
  Phys.\ Lett.\ B {\bf 345}, 414 (1995).

\bibitem{smooth}
 G.~Lazarides and C.~Panagiotakopoulos,
  Phys.\ Rev.\ D {\bf 52} (1995) 559.

\bibitem{ourmut}
  D.~H.~Lyth and E.~D.~Stewart,
  Phys.\ Rev.\ D {\bf 54}, 7186 (1996)

\bibitem{rss}
  M.~ur Rehman, V.~N.~Senoguz and Q.~Shafi,
  Phys.\ Rev.\  D {\bf 75} (2007) 043522
  [arXiv:hep-ph/0612023].

\bibitem{collbrane}
  G.~R.~Dvali and S.~H.~H.~Tye,
  Phys.\ Lett.\ B {\bf 450}, 72 (1999).

\bibitem{andreihybrid}
 A.~D.~Linde,
Phys.\ Rev.\ D {\bf 49} (1994) 748.

\bibitem{brane}
 S.~H.~Henry Tye,
  arXiv:hep-th/0610221.

\bibitem{pseudonatural}
 N.~Arkani-Hamed, H.~C.~Cheng, P.~Creminelli and L.~Randall,
  JCAP {\bf 0307}, 003 (2003)
  [arXiv:hep-th/0302034].

\bibitem{dineriotto}
  M.~Dine and A.~Riotto,
  Phys.\ Rev.\ Lett.\  {\bf 79}, 2632 (1997)
  [arXiv:hep-ph/9705386].

\bibitem{supernew}
 K.~Kumekawa, T.~Moroi and T.~Yanagida,
  Prog.\ Theor.\ Phys.\  {\bf 92} (1994) 437
  [arXiv:hep-ph/9405337];
 K.~I.~Izawa and T.~Yanagida,
  Phys.\ Lett.\  B {\bf 393} (1997) 331
  [arXiv:hep-ph/9608359];
 K.~I.~Izawa, M.~Kawasaki and T.~Yanagida,
  Phys.\ Lett.\  B {\bf 411} (1997) 249
  [arXiv:hep-ph/9707201];
  K.~I.~Izawa,
  Phys.\ Lett.\  B {\bf 576} (2003) 1
  [arXiv:hep-ph/0305286];
  M.~Ibe, K.~I.~Izawa, Y.~Shinbara and T.~T.~Yanagida,
  Phys.\ Lett.\  B {\bf 637} (2006) 21
  [arXiv:hep-ph/0602192].

\bibitem{supernatural}
 L.~Randall, M.~Soljacic and A.~H.~Guth,
  Nucl.\ Phys.\  B {\bf 472}, 377 (1996)
  [arXiv:hep-ph/9512439].

\bibitem{running}
  E.~D.~Stewart,
  Phys.\ Lett.\  B {\bf 391}, 34 (1997)
  [arXiv:hep-ph/9606241];
 E.~D.~Stewart,
  Phys.\ Rev.\  D {\bf 56}, 2019 (1997)
  [arXiv:hep-ph/9703232];
 L.~Covi and D.~H.~Lyth,
  Phys.\ Rev.\  D {\bf 59}, 063515 (1999)
  [arXiv:hep-ph/9809562].

\bibitem{lgl}
  S.~M.~Leach, I.~J.~Grivell and A.~R.~Liddle,
  Phys.\ Rev.\  D {\bf 62}, 043516 (2000)
  [arXiv:astro-ph/0004296].

\bibitem{aterm}
  R.~Allahverdi, K.~Enqvist, J.~Garcia-Bellido and A.~Mazumdar,
  Phys.\ Rev.\ Lett.\  {\bf 97}, 191304 (2006)
  [arXiv:hep-ph/0605035];
R.~Allahverdi, K.~Enqvist, J.~Garcia-Bellido, A.~Jokinen and A.~Mazumdar,
  JCAP {\bf 0706}, 019 (2007)
  [arXiv:hep-ph/0610134].


\bibitem{bdl}
J.~C.~Bueno Sanchez, K.~Dimopoulos and D.~H.~Lyth,
  JCAP {\bf 0701}, 015 (2007)
  [arXiv:hep-ph/0608299].

\bibitem{recentmod}
 D.~Baumann, A.~Dymarsky, I.~R.~Klebanov, L.~McAllister and P.~J.~Steinhardt,
  arXiv:0705.3837 [hep-th];
D.~Baumann, A.~Dymarsky, I.~R.~Klebanov and L.~McAllister,
  arXiv:0706.0360 [hep-th];
  A.~Krause and E.~Pajer,
  arXiv:0705.4682 [hep-th].

\end{thebibliography}
\end{document}